\documentclass[
reprint,           % journal's actual layout
superscriptaddress,% authors with affiliations via superscripts
amsmath,           % add AMS-Latex features
amssymb,           % add extra AMS symbols, including amsfonts
aps,               % aps or aip
prl,               % prl, pra, prb, prc, prd, pre, prstab
notitlepage,       % control appearance of title page
floatfix,          % process floats as early as possible
nofootinbib,
]{revtex4-1}

\usepackage{tensor}     % manipulate tensors
\usepackage{graphicx}   % include figures
\usepackage[
colorlinks=true,        % color link
citecolor=blue,         % cite color
linkcolor=blue,         % link color
urlcolor=blue           % url color
]{hyperref}             % create hyperlinks
\usepackage{bm}         % \bm{<text>} Bold math symbols
\usepackage{xcolor}     % textcolor
\usepackage[flushleft]{threeparttable}
\usepackage{tabulary}
\usepackage{color}      % \textcolor{declared-color}{text}
\usepackage[utf8]{inputenc} % accented characters for .bib file using XeLaTex
\usepackage[section]{placeins} % figures processing
\usepackage{url}
\usepackage{blindtext}
\usepackage[T1]{fontenc}
\usepackage[utf8]{inputenc}

\newcommand{\nc}{\newcommand*} 

\nc{\al}{\alpha}
\nc{\s}{\sigma}
\nc{\dt}{\delta}
\nc{\Dt}{\Delta}
\nc{\Ld}{\Lambda}
\nc{\p}{\partial}
\nc{\om}{\omega}
\nc{\Om}{\Omega}
\nc{\rd}{\mathrm{d}}
\nc{\Od}[1]{\mathcal{O}(#1)} % order operator
\nc{\kp}{\kappa}
\nc{\one}{\uppercase\expandafter{\romannumeral1}}
\nc{\two}{\uppercase\expandafter{\romannumeral2}}
\nc{\three}{\uppercase\expandafter{\romannumeral3}}
\def\({\left(}
\def\){\right)}
\def\[{\left[}
\def\]{\right]}
\def\e{\begin{equation}}
\def\q{\end{equation}}
\def\m{\begin{eqnarray}}
\def\n{\end{eqnarray}}
\nc{\Eq}[1]{Eq.~\eqref{#1}}     % equation
\nc{\Fig}[1]{Fig.~\ref{#1}}     % figure
\nc{\Table}[1]{Table~\ref{#1}}  % table
\nc{\Sec}[1]{Sec.~\ref{#1}}     % section
\nc{\Msun}{M_\odot}             % solar mass
\nc{\fpbh}{f_{\mathrm{pbh}}}    % f_pbh
\nc{\fpbhn}{f_{\mathrm{pbh0}}}    % f_pbh
\nc{\mR}{\mathcal{R}} % merger rate density
\nc{\seq}{\sigma_{\mathrm{eq}}}
\nc{\ogw}{\Omega_{\mathrm{GW}}}
\nc{\gpcyr}{\mathrm{Gpc}^{-3}\,\mathrm{yr}^{-1}}
\nc{\lvc}{LIGO/Virgo} % LIGO-VIRGO collaboration
\nc{\SNR}{\mathrm{SNR}} % signal to noise ratio
\nc{\mmin}{{m_{\mathrm{min}}}}
\nc{\mmax}{{m_{\mathrm{max}}}}
\nc{\Mmin}{{M_{\mathrm{min}}}}
\nc{\fmin}{{f_{\mathrm{min}}}}
\nc{\VT}{\mathrm{VT}}
\nc{\rhoGW}{\rho_{\mathrm{GW}}}
\nc{\vth}{\vec{\theta}}
\nc{\vd}{\vec{d}}
\nc{\vla}{\vec{\lambda}}
\nc{\Nobs}{N_{\mathrm{obs}}}
\nc{\av}[1]{\langle #1 \rangle} % average bracket
\nc{\km}{\mathrm{km}}
\nc{\Mpc}{\mathrm{Mpc}}
\nc{\Tobs}{T_{\mathrm{obs}}}
\nc{\Ntemp}{N_{\mathrm{temp}}}
\nc{\addref}{[\textcolor{red}{add ref}] } % placeholder of references
\nc{\eg}{\textit{e.g.~}}
\nc{\app}{\approx}
\nc{\hf}{\frac{1}{2}}
\nc{\discuss}{\textcolor{red}{Add discussion here!}}
\nc{\red}[1]{\textcolor{red}{#1}}
\nc{\mH}{\mathcal{H}}
\nc{\cs}{c_s^2}
\nc{\Sij}[1]{S_{ij}^{(#1)}}
\nc{\vi}[1]{v_i^{(#1)}}
\nc{\no}{\nonumber}
\def\<{\left\langle}
\def\>{\right\rangle}

\nc{\bk}{\bm{k}}
\nc{\bq}{\bm{q}}
\nc{\bp}{\bm{p}}
\nc{\bl}{\bm{l}}
\nc{\bx}{\bm{x}}
\nc{\be}{\mathbf{e}}
\nc{\mS}{\mathcal{S}}
\nc{\te}{\tilde{\eta}}
\nc{\tp}{\tilde{p}}
\nc{\tk}{\tilde{k}}
\nc{\tx}{\tilde{x}}
\nc{\tF}{\tilde{F}}
\nc{\tA}{\tilde{A}}
\nc{\mkpq}{|\bk-\bp-\bq|}
\nc{\mpq}{|\bp-\bq|}
\nc{\mkp}{|\bk-\bp|}
\nc{\mSi}[1]{\mS^{(#1)}({\bk, \eta})}
\nc{\vk}{\vec{k}}
\nc{\kstar}{k_*}
\nc{\xstar}{x_*}
\nc{\mpbh}{m_{\rm{pbh}}}
\renewcommand{\vec}[1]{\boldsymbol{#1}} % Uncomment for BOLD vectors.

\begin{document}
%%%%%%%%%%%%%%%%%%%%%%%%%%%%%%%%%%%%%%%%%%%%%%%%%%%%%%%%%%%%%%%%%%%%%%%%%%%%%%%%
	
%%%%%%%%%%%%%%%%%%%%%%%%%%%%%%%%%%%% title %%%%%%%%%%%%%%%%%%%%%%%%%%%%%%%%%%%%%
\title{Stellar rotation as a new observable to test general relativity in the Galactic Center}

\author{Yun Fang}
\email{fang.yun@pku.edu.cn}
\affiliation{Kavli Institute for Astronomy and Astrophysics at Peking University, Beijing 100871, China}
%%%%%%%%%%%%%%%%%%%%%%%%%%%%%%%%%%%%%%%%%%%%%%%%%%%%%%%%%%%%%%%%
\author{Xian Chen}
\email{xian.chen@pku.edu.cn}
\affiliation{Astronomy Department, School of Physics, Peking University, Beijing 100871, China}
\affiliation{Kavli Institute for Astronomy and Astrophysics at Peking University, Beijing 100871, China}
%%%%%%%%%%%%%%%%%%%%%%%%%%%%%% author 2 %%%%%%%%%%%%%%%%%%%%%%%%%%%%%

%%%%%%%%%%%%%%%%%%%%%%%%%%%%%%%%%%%%% date %%%%%%%%%%%%%%%%%%%%%%%%%%%%%%%%%%%%%
\date{\today}

%%%%%%%%%%%%%%%%%%%%%%%%%%%%%%%%% abstract %%%%%%%%%%%%%%%%%%%%%%%%%%%%%%%%%%%%%
\begin{abstract}
S-stars in the Galactic Center are excellent testbeds of various general
relativistic effects. While previous works focus on modeling their orbital motion
around Sgr A*--the supermassive black hole in the Galactic Center--here
we explore the possibility of using the rotation of S-stars to test the
de Sitter precession predicted by general relativity.  We show that by reorienting the 
rotation axes of S-stars, de Sitter precession will change the
apparent width of the absorption lines in the stellar spectra.
Our numerical simulations suggest that the newly discovered S4714
and S62 are best suited for such a test because of their small
pericenter distances relative to Sgr A*. Depending on the initial
inclination of the star, the line width would vary by as
much as $20-76\,{\rm km\,s^{-1}}$ within a period of $20-40$ years.
Such a variation is comparable to the current detection limit.  Since
the precession rate is sensitive to the orbital eccentricity and
stellar quadrupole structure, monitoring the rotation of S-stars could also help us better constrain the
orbital elements of the S-stars and their internal structures.
\end{abstract}

\pacs{???}
	
\maketitle

\section{introduction}

The predictions of general relativity (GR) have been tested in a variety of
astrophysical systems.  They are tested with high precision in our solar
system, e.g., the advance of the perihelion of Mercury (``Schwarzschild
precession'') \cite{Will1993book}, the ``de Sitter'' (or geodetic) precession 
in the earth-moon system due to the presence of the sun \cite{ShapiroPRL1988, WILLIAMS2002}, 
and the
Lense-Thrring precession of a man-made gyroscope orbiting around the earth
\cite{Ciufolini2004}.  GR also has passed test in the region of a much stronger
field,  such as in several pulsar systems \cite{Kramer2006science,
Kramer:2016kwa}, close to the supermassive black hole (SMBH) in M87 by imaging
the shadow \cite{Akiyama:2019cqa}, as well as in the mergers of compact objects
through detecting their gravitational waves \cite{Abbott:2016blz, LIGO_GWTC1,
LIGO_GWTC2}.  

The centers of galaxies are natural laboratories of strong gravity because they
usually contain SMBHs \cite{Ghez2009astro, Kormendy_2013, McConnell2013ApJ}.  In fact, the
Galactic Center also harbors a SMBH, known as Sagittarius A* (Sgr A*), which
has a mass of $4 \times 10^6 \Msun$ \cite{Ghez2008ApJ, GenzelRMP2010}.  It is
surrounded by a dense cluster of young stars, called S-stars.  Several of them,
such as the famous S2 star, are orbiting Sgr A* with such a small distance and
high eccentricity that the velocity reaches a few percent of the speed of light
when the star passes the pericenter \cite{Ott2002Nature,
Meyer2012Sci,Boehle_2016,Peisker2020ApJ}.  These stars are excellent testbeds
of various relativistic effects. Using S2, we have detected the Schwarzschild
precession \cite{Abuter:2020dou} and  the gravitational redshift of the
spectral lines \cite{Abuter:2018drb, DoTuan2019}.  Continuous monitoring the
orbits of S-stars would further allow us to test the ``no hair theorem''
\cite{Will_2008APJL, Merritt2010PhRvD}, constraining the possible fifth force
of the unification theory \cite{Hees:2017aal}, probing the hidden companion of
Sgr A* if there is one \cite{Naoz_2019}, and constraining the properties of the
dark matter around the SMBH \cite{Becerra-Vergara:2020xoj}. 

One essential component that is still missing in the current GR experiment
using S-stars is stellar rotation. GR predicts that a rotating star, like a
gyroscope, will undergo de Sitter precession around a SMBH.  Just like the
Schwarzschild precession, de Sitter precession is also a first-order
post-Newtonian effect so that it becomes significant when the S-star passes the
pericenter of its orbit around Sgr A*.  It has been suggested that the de
Sitter precession could be detected using a pulsar with an orbital period
shorter than $100$ years around Sgr A* \cite{Pfahl_2004}, but so far no such
pulsar has been discovered.

Two recent discoveries motivated us to study the de Sitter precession of
S-stars. First, a few faint stellar objects were found on orbits closer than S2
relative to Sgr A* \cite{Peisker2020ApJ}.  Their orbital eccentricities are
also very large, making the pericenter distances much closer to Sgr A* as well.
Second, the same observation revealed that at least one of the newly found
S-stars, S4711, is fast rotating at a projected velocity of $V\sin i=239.60 \pm
25.21\, {\rm km\,s^{-1}}$.  This projected rotation velocity is measured from
the width of the absorption lines in the stellar spectrum.  This result, as
well as the earlier observation that S2  has a projected rotation velocity of 
$100 \pm$ $30\,{\rm km\,s^{-1}}$ \cite{Martins:2007rv}, suggest that many, if
not all, S-stars are fast rotators.  Since a rotating star around Sgr A* should
undergo de Sitter precession, we would expect the angle $i$ between the
rotation axis and the line-of-sight to change whenever the star passes by the
orbital perecenter.  Such a change will induce a variation in the apparent line
width, which could serve as a new observable for us to test GR in the Galactic
Center.

\section{Theory}

The precession law of a rotating star with a spin angular momentum {\bf J} revolving around a SMBH is governed by the equation 
\begin{eqnarray}\label{precession}
  {d {\bf J} \over dt}=( {\bf \Omega}^{(\text{geod})}+ {\bf \Omega}^{(\text{LT})}) \times {\bf J} + {\bf N}^{\text{quad}} 
 \end{eqnarray}
(see, e.g. \cite{Thorne:1984mz}),
where ${\bf \Omega}^{(\text{geod})}$ is  the rate of
geodetic precession determined by the mass $M$ of the SMBH, ${\bf
\Omega}^{(\text{LT})}$ is the Lense-Thirring precession rate determined by the
spin angular momentum ${\bf S}$ of the SMBH, and ${\bf N}^{\text{quad}}$ is the
torque induced by the interaction between the mass quadrupole moment of the star and 
the tidal force of the SMBH.

To facilitate the calculation, we define ${\bf r}$ as the relative distance
from the SMBH to the star, so that ${\bf v}=\dot{\bf r}$ is the orbital
velocity of the star and ${\bf n}:={\bf r}/r$ is the direction vector. With
these terms, we can write
\begin{eqnarray}
{\bf \Omega}^{(\text{geod})}={3\over 2 } {M\over r^2} {\bf v}\times {\bf n}. 
\end{eqnarray}
We have assumed $G=c=1$ in the above calculation.
The precession rate due to the Lense-Thirring effect can be calculated with
\begin{eqnarray}
{\bf \Omega}^{(\text{LT})}={1\over r^3} [ -{\bf S}+3 {\bf n} ({\bf n}\cdot {\bf S}) ], 
\end{eqnarray}
with is of order ${\cal O}(v)$ smaller than the geodetic precession 
and hence neglected in the later analysis. 

The last term ${\bf N^{\text{quad}}}$ in Equation~(\ref{precession}) is 
related to the stellar mass quadrupole moment tensor $\mathop{{\bf Q}}$ as
\begin{eqnarray}
{\bf N}^{\text{quad}}=- 3 {M\over r^3}{\bf n}\times \mathop{{\bf Q}} \limits ^{\leftrightarrow} \cdot {\bf n},
\label{define_N}
\end{eqnarray}
and he components of $\mathop{{\bf Q}}$ is given by
\begin{eqnarray}
Q^{jk}=\left [ \int \rho x^j x^k d^3 x \right ] ^\text{STF}, 
\end{eqnarray}
where $\rho$ denotes the density of the star and the superscript
``$\text{STF}$'' means the symmetric and trace-free part. 

To  be  self-consistent,  we  assume  that  the  non-vanishing part of the mass quadrupole moment is induced by the rotation of the star. 
Under the symmetry condition and in the
coordinate system $(x', y', z')$ where $z'$ is aligned with the ${\bf J}$ axis, 
the tensor $\mathop{{\bf Q}}$ takes a simple form in which
$I_{x'y'}=I_{x'z'}=I_{y'z'}=0$ and $I_{x'x'}=I_{y'y'}\neq I_{z'z'}$. 
It follows that
\begin{eqnarray}
{Q}^{i'j'}=(I_{x'x'}-I_{z'z'})
\left( \begin{matrix} 
    {1 \over 3}  &   0 & 0  \\   
    0   &   {1\over 3}  &  0\\   
    0   &  0   &  -{2\over 3}
\end{matrix} \right). 
\end{eqnarray}
To further derive the component of $\mathop{{\bf Q}}$ in an arbitrary coordinate system
$(x, y, z)$, we use the transformation matrix
\begin{eqnarray}
R(i, \kappa)= 
\left(
\begin{array}{ccc}
 \cos \kappa  & \sin \kappa  & 0 \\
 -\sin \kappa  & \cos \kappa  & 0 \\
 0 & 0 & 1 \\
\end{array}
\right)
\left(
\begin{array}{ccc}
 1 & 0 & 0 \\
 0 & \cos i  & \sin i  \\
 0 & -\sin i  & \cos i  \\
\end{array}
\right),
\end{eqnarray}
where $i$ is the angle between the $z$ and $z'$-axis, $\kappa$ is the angle
between the $y$-axis and the projection of the $z'$-axis in the $x-y$
plane, thus $(i, {\pi \over 2}-\kappa)$ is the spherical coordinate of the
$z'$-axis in the $(x, y, z)$ frame. Then we can derive $Q^{ij}$ in the $(x, y,
z)$ frame using the transformation $Q=R Q' R^{\text{T}}$.

\section{Connecting theory with observation}

If the direction of ${\bf J}$ varies, the rotation velocity projected along the
line-of-sight, $V\sin i$, will change accordingly.  The projected velocity
$V\sin i$ is related to the observed width $\Delta\lambda$ of the absorption
lines in the spectra of S-stars. The relationship is $\Delta \lambda= 2 \lambda
V \sin i$, where $\lambda$ is the wavelength of the centroid of an spectral
line. Therefore, we can use the variation of $\Delta\lambda$ to test the
prediction of Equation~(\ref{precession}).

\begin{figure}%[h!] %H
\centering 
\includegraphics[height=5cm]{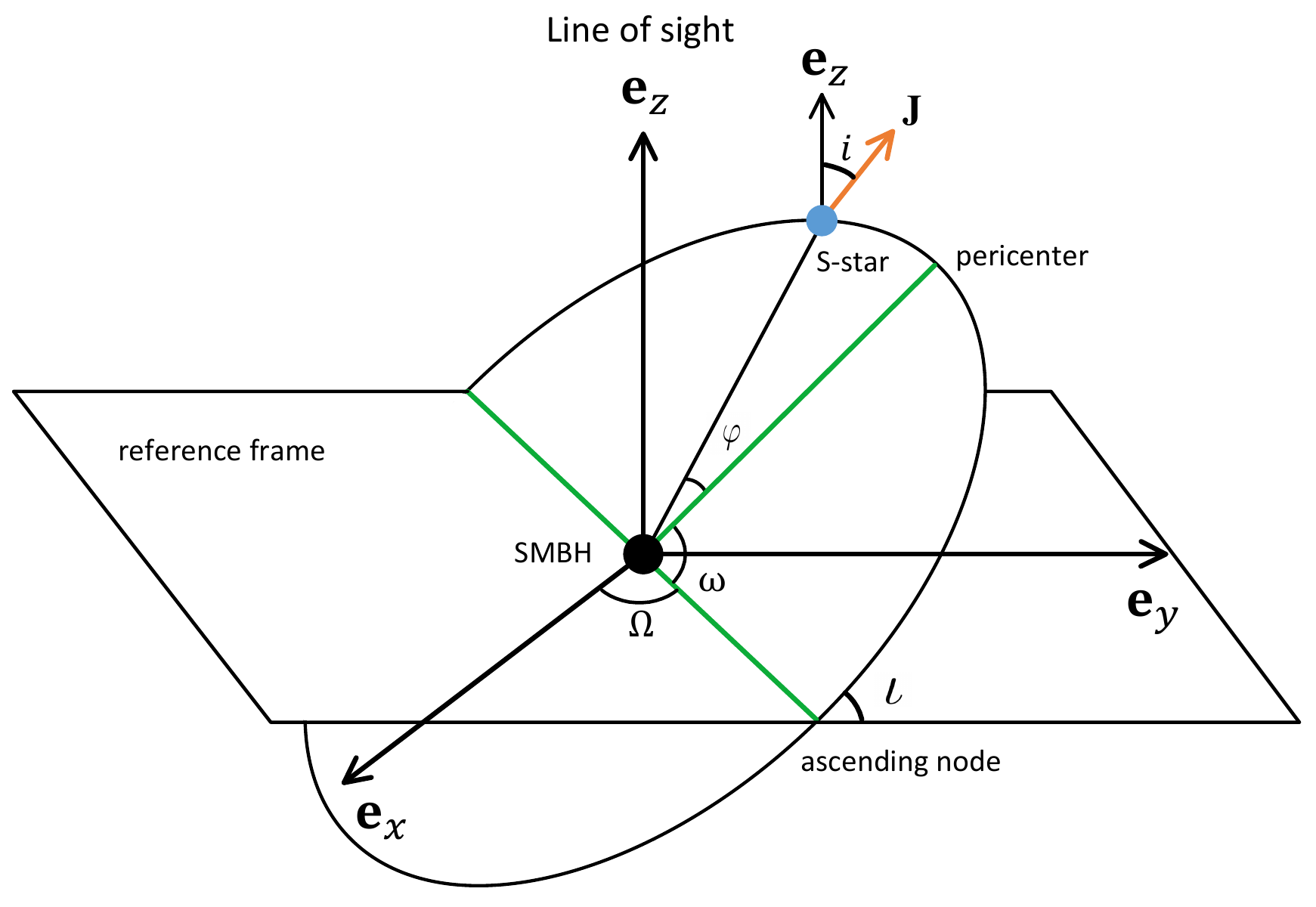}  
\caption{Definition of the orbital elements of an S-star (blue dot) in a
reference frame centered on the SMBH (black dot) and with the $z$-axis
aligned with the line-of-sight (observer resides in the $-z$ direction). 
The orange arrow shows the rotating
axis, i.e., the direction of the spin angular momentum, of the star.
} 
\label{Figure.orbit} 
\end{figure}

In the following, we simulate the evolution of $V\sin i$ for four S-stars,
namely S4711, S4714, S62, and S2, by numerically integrating
Equation~(\ref{precession}). S2 is the most famous one due to its early
detection, closeness to Sgr A*, and large eccentricity. It has been monitored
for almost three decades by SINFONI, NACO, and, more recently, GRAVITY
\cite{Abuter:2018drb, Abuter:2020dou}. The orbital period is about $16$ years
and the eccentricity is about $0.88$.  The three newly discovered S-stars,
S4711, S4714 and S62, have even shorter periods than S2 as well as relatively
large orbital eccentricities \cite{Peisker2020ApJ}.  These properties make them
ideal test subjects of the de Sitter precession.  The orbital parameters of our
chosen S-stars are given in Table~\ref{table:stars}, and the meanings of the
parameters are illustrated in Figure~\ref{Figure.orbit}.

\begin{table*}[]
\begin{threeparttable}
\caption{The orbital parameters and simulation results for S2, S4711, S4714, and S62.}
    \centering
    \begin{tabular}{cccccccccc}
            \hline
            \hline
            name & $m_*$ ($\Msun$) & $a$ (mpc) & $e$ & $\iota$ ($^\circ$) & $\omega$ ($^\circ$) & $\Omega$ ($^\circ$)&  $t_{\rm period}$ (years) & $\Delta V (q=0)$ & $\Delta V (q=0.1)$ \\
            \hline
            %\hline
             %
             S2 & 10 & 4.895 & 0.886 & 133.9 & 66.0 & 227.4 &  16 &  $0.4\,{\rm km\,s^{-1}}$ & $0.6\,{\rm km\,s^{-1}}$ \\
             S4711 & 2.2 & 3.002 & 0.768 & 114.71 & 131.59 & 20.10 & 7.6 & $0.4\,{\rm km\,s^{-1}}$ & $0.4\,{\rm km\,s^{-1}}$ \\
            S4714 & 2.0 & 4.079 & 0.985\! $\pm$\! 0.011 & 127.70 & 357.25 & 129.28 &  12.0 &  $53\,{\rm km\,s^{-1}}$ & $76\,{\rm km\,s^{-1}}$ \\
             S62 & 6.1 & 3.588 &  0.976\!$\pm$\! 0.01 & 72.76 & 42.62 & 122.61 &  9.9  & $21\,{\rm km\,s^{-1}}$ & $25\,{\rm km\,s^{-1}}$ \\ 
            \hline
    \end{tabular}
     \begin{tablenotes}
     \item {\bf Notes}: The orbital parameters of S2 are from
\cite{Martins:2007rv} and \cite{DoTuan2019}, and those of S4711, S4714, and S62 are from \cite{Peisker2020ApJ}. 
For each star we adopt from the above references the mass $m_*$, semi-major axis $a$, inclination $\iota$, phase of pericenter $\omega$, angle of ascending node
$\Omega$, and orbital period $t_{\rm period}$.   The last two columns $\Delta V(q=0)$ and $\Delta V(q=0.1)$ are the maximum variation of the line width, $V \sin{i}$, within a period of $40$ years computed using two different stellar ellipticities, $q=0$ and $q=0.1$. 
\end{tablenotes} \label{table:stars} 
\end{threeparttable} 
\end{table*}

We set the mass of the SMBH to be
$M=4 \times 10^6\Msun$. We assume that the stellar rotation velocity in the
equatorial plane is $V=300\,{\rm km\,s^{-1}}$, to be consistent with the
detected velocity in \cite{Peisker2020ApJ}. We note that
changing the initial $V$ will not affect the evolution of $i$.
Therefore, if we adopt a lower rotation velocity,
the projected line width and its variation with time would be proportionally smaller.

The quadrupole of the star is
computed with
\begin{equation}
  I_{x'x'}=I_{y'y'}=(1+q) I_{z'z'}=\frac{\epsilon m_* {r_*}^2}
{\sqrt{2+1/(1+q)^2}} ,
\end{equation} 
where $m_*$ is the mass of the star, $r_*$ is the stellar radius, $q$ is the
ellipticity of the star, and $\epsilon$ is a parameter accounting for the
internal mass distribution of the star.  Moreover, we compute the spin angular momentum
of the star with ${\bf{J}}=\epsilon_J m_* r_* V {\bf j}$, where ${\bf j}:={\bf
J}/|{\bf J}|$ is the direction of ${\bf J}$ and $\epsilon_J$ is another
dimensionless parameter characterizing the internal rotating structure of the
star.

We note that the result of the numerical integration is insensitive to the
internal structure of the star but mainly depends on the size $r_*$ and the
ellipticity $q$. The reasons are as follows.  First, the stellar mass $m_*$
cancels out on both sides of Equation~(\ref{precession}) because both $\bf{J}$
and ${\bf N}^{\text{quad}}$ are proportional to $m_*$. Second, the parameter
$\epsilon$ in ${\bf N}^{\text{quad}}$ and the parameter $\epsilon_J$ in
$\bf{J}$ also cancel out if the star rotates as a rigid body, because in this
case both parameters scale with $\int\rho r'^4 dr'$. For non-rigid-body
rotation, $\epsilon$ and $\epsilon_J$ would be different, which would result in
a different relative contribution to the precession rate by the two terms on
the right-hand-side of Equation~(\ref{precession}). However, numerically, the
net effect can be absorbed into the uncertainty of the value of $q$.

We adopt a value of  $r_*=10\,R_\odot$ for S2 \cite{Martins:2007rv}. For the
other three S-stars, we derive their radii from the mass-radius relationship
presented in \cite{Tout1996mnras}. For the ellipticity $q$, since there is
currently no observational constraint, we take $0$ and $0.1$ as two
representative values. The latter value is predicted by the model of
early-type stars \cite{Gagnier2019A}.

\section{Results}

\begin{figure}%[t] %H 
\centering 
\includegraphics[height=11cm]{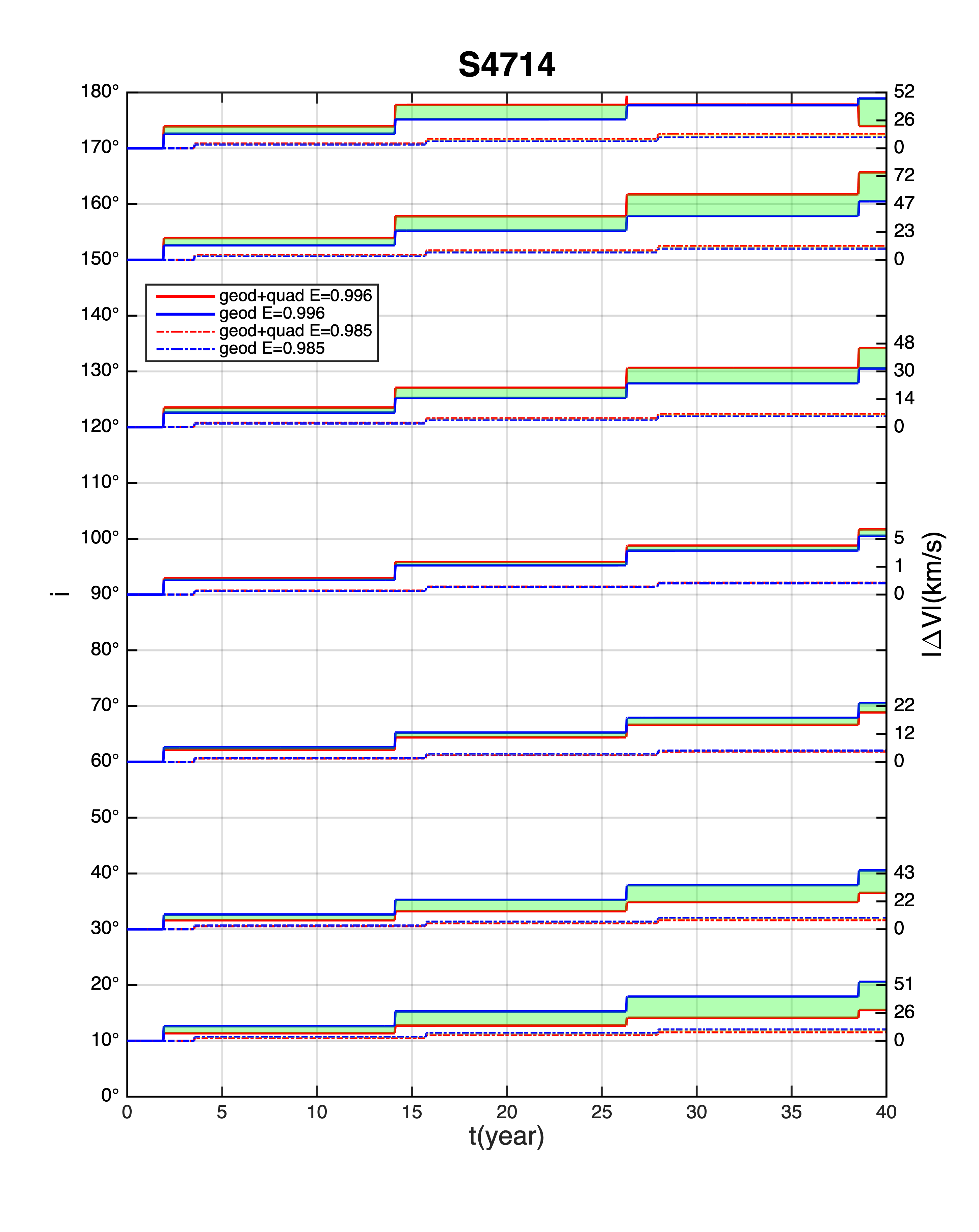}
\caption{Evolution of the inclination angle (left ordinate) for S4714, as well
as the corresponding variation of the apparent (projected) rotation velocity
(right ordinate). We tried seven different initial values for the inclination
angle, while the initial azimuth angle of the rotation axis ($\bf{j}$) is fixed
to $120^{\circ}$. The dot-dashed lines are the results derived using the medium
value $e=0.985$ of the orbital eccentricity, and the solid lines are derived
using the maximum eccentricity $e=0.996$ allowed by observation within the
uncertainty. The red lines are computed assuming that the quadrupole momentum
of the star is zero and the blue ones are assuming $q=0.1$. Therefore, the
shaded area between the red and blue lines indicates the uncertainty due to the
unknown stellar structure. The sudden jumps in the evolution of $i$ coincide with
the star passing by the pericenter, where both the de Sitter precession and the 
mass-quadrupole effect become the most significant. } \label{Figure.S4714} 
\end{figure}

{\bf S4714} has the smallest pericenter distance and hence is best suited for the
test of the de Sitter precession. Figure~\ref{Figure.S4714} shows the evolution
of the orientation of ${\bf J}$ within $40$ years.  We tried seven different
initial conditions for the inclination angle, namely, $i_0=10^{\circ}$,
$30^{\circ}$, $60^{\circ}$, $90^{\circ}$, $120^{\circ}$, $150^{\circ}$, and
$170^{\circ}$. The right ordinate corresponds to the variation of the apparent
(projected) rotation velocity, $|\Delta V|$, which is defined as $|V\sin
i-V\sin i_0|$. We also allowed the orbital eccentricity $e$ to vary between the
medium and maximum values allowed by observation, and the stellar ellipticity
$q$ to vary between $0$ and $0.1$, to accommodate the observational and
theoretical uncertainties.

Figure~\ref{Figure.S4714} shows that larger orbital eccentricity in general
leads to a larger variation of the projected rotation velocity $|\Delta V|$.
Moreover, the quadrupole moment of the star suppresses the precession of the
stellar rotation axis when $i<90^\circ$ and enhances the precession when
$i\ge90^\circ$. These results indicate that measuring $|\Delta V|$ would enable
us to better constrain the orbital elements as well as probe the internal
structure of S4714.

Regarding detectability, if $e=0.985$, i.e., the medium value given by
observation, we find that $|\Delta V|$ remains smaller than the current
detection limit, $25\,{\rm km\,s^{-1}}$ \cite{Peisker2020ApJ}, during the $40$
years of evolution.  However, when $e$ increases to the maximum value allowed
by observation, i.e., $e=0.996$, $|\Delta V|$ could exceed $25\,{\rm
km\,s^{-1}}$ within $20$ years if the initial inclination angle is
$i_0=10^\circ$, $30^\circ$, $150^\circ$, or $170^\circ$.  If the observational
period could be elongated to $40$ years, an initial inclination angle of
$i_0=120^\circ$ would also lead to an detectable $|\Delta V|$. Moreover, if in
the future the detection limit of the rotation velocity could be improved to
$20\,{\rm km\,s^{-1}}$, the variation $|\Delta V|$ induced by an initial
inclination angle of $i_0=60^\circ$ would also become detectable within a
period of $40$ years. To detect the $|\Delta V|$ corresponding to an initial
inclination of $i_0=90^\circ$, an observational period of $40$ years and a
detection limit of $5\,{\rm km\,s^{-1}}$ are preferred.  For easier comparison
with other stars, the maximum $|\Delta V|$ from our numerical simulations are
given in the last two columns of Table~\ref{table:stars}.

\begin{figure}%[t] %H
\centering 
\includegraphics[height=11cm]{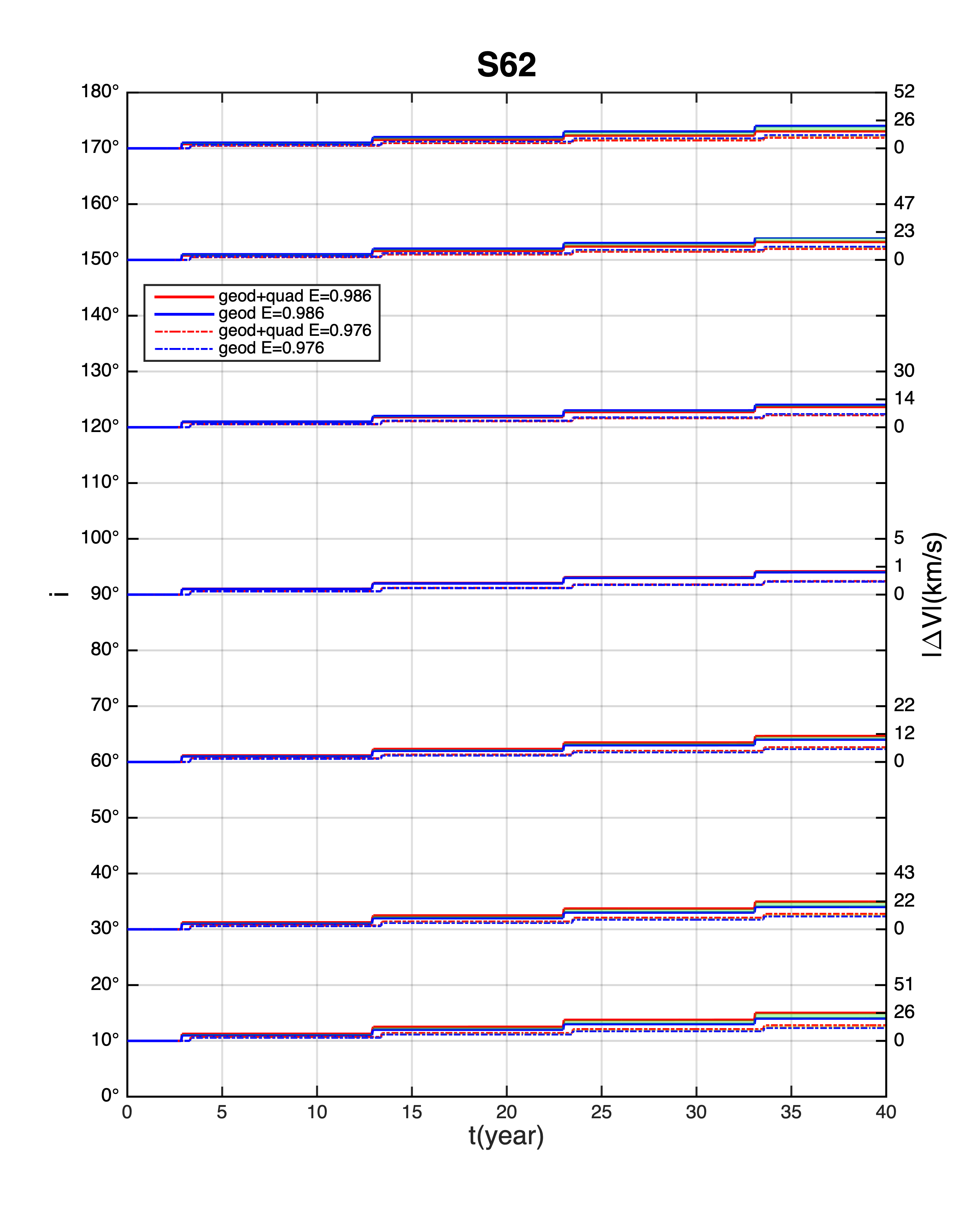}  
\caption{The same as Fig.~\ref{Figure.S4714} but for S62. 
} 
\label{Figure.S62} 
\end{figure}

{\bf S62} also has a relatively small pericenter distance.  The simulation results
for this star are shown in Figure~\ref{Figure.S62}.  Comparing with the
previous results for S4714, we find two differences.  First, the variation of
the projected rotation velocity is much smaller than that of S4714.  Second,
the lines taking into account the quadrupole moment of the star (red) are less
separated from the line without the quadrupole moment (blue).  Both differences
are caused by the larger pericenter distance of S62 relative to that of S4714.
Because of the larger pericenter distance, the magnitude of $\bf{\Omega}^{\rm
geod}$, which determines the rate of de Sitter precession, and the ratio
$|\bf{N}^{\rm quad}|/|\bf{\Omega}^{\rm geod}\times\bf{J}|$, which determines
the relative importance of the quadrupole moment, both become smaller.
Nevertheless, when we set $e=0.986$ (solid lines), i.e., the maximum eccentricity allowed by
observation, the variation of $V \sin{i}$ within $40$ years is
comparable to the current detection limit of $25\,{\rm km\,s^{-1}}$ if
$i_0=10^\circ$ or $170^\circ$. The maximum values of $|\Delta V|$
are also given in Table~\ref{table:stars}.

{\bf S2} and {\bf S4711} have larger pericenter distances than those of S4714
and S16. For this reason, their $|\Delta V|$ is too small to be detected.
We do not show the evolution of their inclination angles but only give the
maximum $|\Delta V|$ derived from our numerical simulations in Table~\ref{table:stars}.

%%%%%%%%%%%%%%%%%%%%%%%%%%%%%%%%%%%%%%%%%%%%%%%%%%%%%%%%%%%%%%%%%%%%%%%%%%%%%%

\section{Conclusion}

In this Letter, we pointed out that the de Sitter precession induced by Sgr A*
on the surrounding S-stars could be detected through the variation of the
width of the spectral lines. Our numerical
simulations suggested that the newly discovered S4717 and S62 are best suited for
such a test because of their small pericenter distances relative to Sgr A*.  We
showed that within a period of $20-40$ years, their line width would vary by
as much as $20-76,{\rm km\,s^{-1}}$, depending on the initial inclination of the
rotation axes. Such a variation is comparable to the current detection limit.
Since the precession rate is sensitive to the eccentricity of the stellar orbit
as well as the quadrupole moment of the star, observing the variation of the line width
would also help us better constrain the orbital elements and probe the internal
structures of the S-stars.

	%%%%%%%%%%%%%%%%%%%%%%%%%%%%%%%% acknowledgments %%%%%%%%%%%%%%%%%%%%%%%%%%%%%%%

This work is supported by the National Science Foundation of China (NSFC) grant
No 11721303. XC acknowledges the support by NSFC grants No 11873022, and
11991053. 
	%%%%%%%%%%%%%%%%%%%%%%%%%%%%%%%%%%%%%%%%%%%%%%%%%%%%%%%%%%%%%%%%%%%%%%%%%%%%%%%%
	%\bibliographystyle{apj}
%	\bibliography{./reference}
	
%%%%%%%%%%%%%%%%%%%%%%%%%%%%%%%%%%%%%%%%%%%%%%%%%%%%%%%%%%%%%%%%%%%%%%%%%%%%%%

%\nocite{*}
\bibliographystyle{apsrev4-1.bst}
\bibliography{reference}% Produces the bibliography via BibTeX.

\end{document}